# Ar$^+$-sputtered Ge (001) surface nanostructuring at high implant temperature

Debasree Chowdhury[1*] and Debabrata Ghose

*Saha Institute of Nuclear Physics,*

*Sector – I, Block – AF, Bidhan Nagar, Kolkata 700064, India*

**Abstract**

Ion sputtering induced nanoscale pattern formation on Ge (001) surface by 500 eV Ar$^+$ bombardment has been investigated for a wide range of ion incidence angles at temperature of 300$^0$C. A fourfold symmetric topography forms in the angular regime 0 - 65$^0$. Above 65$^0$, they show a remarkable transition into highly regular one-dimensional (1D) asymmetric pattern, known as perpendicular mode ripples. In order to understand growth dynamics of both kind of patterns, we have investigated their temporal evolution as a function of ion fluence in a wide range from $1 \times 10^{17}$ to $1 \times 10^{20}$ ions cm$^{-2}$. In addition, we study the effect of substrate rotation on Ge surface morphology in whole angular range. The four-fold symmetric patterns effect does not found to alter their symmetry, while the ripples degenerate into hole structure with a weak fourfold symmetric pattern. The origin of square topographies and their symmetry independency on ion incident angle in the range 0$^0$ to 65$^0$ can be attributed to the growth process due to biased diffusion of vacancies arising from Ehrlich-Schwoebel barrier. Whereas, the ripple formation at grazing incidence angles indicates the dominance of curvature dependent surface instability induced by the incident ion direction.

Keywords: Ge (001) surface; Ar$^+$ sputtering; checkerboard pattern; nanoripple.

___________________________________

\*Corresponding author.

E-mail address: debu.sinp@gmail.com, dchowdhu@ictp.it

[1]Present address: Dipartimento di Fisica, Università di Genova, via Dodecaneso 33, I-16146 Genova, Italy

# 1. Introduction

Low energy ion beam sputtering (IBS) on solid is known to be a simple technique for producing self-organized structures in the nanometer length scale. It is advantageous for its ability of nanostructuring in large area (order of $cm^2$) within few minutes in just one step without any mask, resist or chemical hazards. Two-dimensional dot and one-dimensional ripple topographies are generally found to develop on the surfaces of semiconductors due to IBS at room temperature [1-6]. Experimentally, the pattern formation has been studied mostly with Si [3-8] because of its easy availability and technological importance in semiconductor industry [9]. Ge is also noteworthy in device fabrication technology [10], instead only a few works of Ge nanostructuring via the one step IBS are reported so far [3, 11-18] where the formation of swelling at normal incidence angle (e.g. [16]) and ripple and faceted pattern at high incidence angles are observed. These experiments were carried out at room temperature and unfortunately, the room temperature IBS of semiconductors causes surface amorphization which significantly limits its electrical conductivity and hence, deteriorates the functionality of the nanoscale devices. However, the crystallinity of the sample can be maintained if the irradiation is performed near or above the recrystallization temperature or can be recovered by post-bombardment annealing of the sample. In the former case, the dynamic annealing of the ion-induced defects keeps the sample crystalline and texturing becomes well-ordered and nearly defect-free. Such crystalline Ge nanostructures exhibit high potential in different activities, ranging from optics to optoelectronics [19], thermoelectrics [20], photovoltaics [21, 22], energy storage devices [23] etc. Another advantage of the high temperature irradiation is the activation of the terrace diffusion barrier known as Ehrlich-Schwoebel (ES) barrier [24, 25] which produces a new kind of surface instability behind the formation of nanopattern. The ES barrier at the step edges causes anisotropic surface diffusion of adatoms and vacancies created by ion impact, which results in the growth of instabilities perpendicular to the step direction and develops structures according to the symmetry and orientation of the crystal face [26].

Recently, the ES barrier induced pattern formation on Ge crystal surface by means of IBS at high temperature attracts considerable attention [27-31]. The remarkable work of Ou et al. [27] on hot Ge by 1 keV $Ar^+$ sputtering illustrates the formation of ES barrier driven dense arrays of crystalline alternating mounds and pits, resembling homoepitaxial growth of checkerboard structure as obtained in molecular beam epitaxy (MBE) [32]. In our earlier work [28], we also demonstrated the similar kind of pattern formation through the $Ar^+$ sputtering near the sputter-threshold energy ~ 30 eV. Since both the investigations were done at normal ion beam incidence, it highly motivates us to extend the pattern formation study at off-normal angle of incidences, especially at grazing incidence angles, in order to realize whether such diffusion-bias-generated instability can influence the morphological evolution at grazing incidence too or some other physical phenomena takes over. For the further understanding of the pattern formation mechanism, one of the approach is to study the evolution dynamics of patterns due to off-normal ion incidences on Ge at elevated temperature which have not receive attention so far.

In this paper, we represent the evolution of nanoscale patterns of Ge (001) for whole angular range $0^0$ to $85^0$ by 500 eV $Ar^+$ IBS at high temperature. We observed four-fold symmetric checkerboard morphology which sustains up to the incidence angle of $65^0$ and after that there is a transition to a two-fold symmetric ripple pattern is observed. The four-fold symmetric pattern remains unaffected when the substrate is rotated during ion irradiation, thus showing the importance of anisotropic diffusion dynamics similar to those involved in the case of MBE. On the other hand, the substrate rotation breaks the two-fold symmetric ripple morphology thereby revealing the dominance of curvature-dependent sputtering mechanism at high incidence angles. We also studied the growth dynamics of both of these structures by examining their temporal evolution as a function of ion fluence ranging from $1 \times 10^{17}$ to $1 \times 10^{20}$ ions cm$^{-2}$. Here, we choose ion energy for irradiation low (~ 500 eV) because of their low penetration depth in solid (~2 nm in Ge according to SRIM 2008 [33]) which in turn yields less amorphization and thus, easy to regain surface crystallinity upon annealing. The importance and usefulness of crystalline patterns from application perspective is already discussed before. Going low beyond 500 eV in our equipment will reduce the ion current density which in consequence will increase the sputtering time a lot to reach the range of high ion fluences (~ $10^{20}$ cm$^{-2}$) chosen in this investigation.

## 3. Experimental

Ion irradiation were carried out on commercially available epi-polished Ge (001) wafers with a beam of diameter 4 cm extracted from an inductively coupled RF discharge ion source (M/s Roth & Rau Microsystems GmbH, Germany) [4]. Before irradiation, the wafers were ultrasonically cleaned in acetone followed by methanol for 5 min in each. During irradiation, the chamber pressure was reduced to $10^{-4}$ mbar, while usually the base pressure remains around $10^{-8}$ mbar. Ion incidence angle was varied from 0-90° with respect to substrate normal. There is also the provision to rotate azimuthally the substrate chuck around its axis at the speed of 5 rpm. A schematic diagram of ion beam irradiation geometry is depicted in fig. 1. The irradiation was performed at $300^0$C which is well above the recrystallization temperature ($T_c$) of Ge ($T_c \approx 270^0$C) [34]. Substrate temperature was raised using a radiation heater mounted at the top of the chamber which can increase the temperature up to 450°C. The surface temperature was measured by a calibrated thermocouple mounted behind the target holder. Care has been taken to prevent surface contamination with metallic impurities from the chamber wall or the ion-

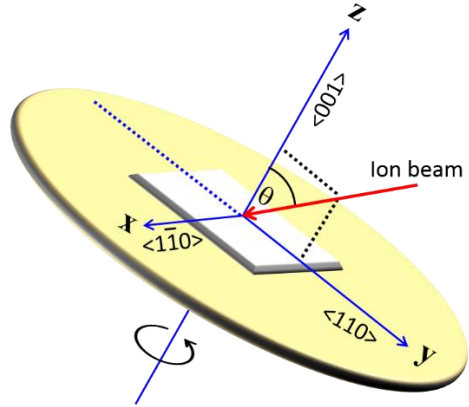

Fig. 1: The schematic diagram represents experimental geometry during IBS. The projection of incidence beam lies parallel to <110> crystallographic direction of the sample surface. The rotation was given around substrate normal ('z' axis).

-source. Ion current density was kept constant around 1000 µA cm$^{-2}$. After irradiation, the surface topography was examined by atomic force microscope (VeccoNanoscope IV multimode microscope) under ambient condition operating in tapping mode with Si cantilevers of nominal tip radius 10 nm. The crystalline quality of the irradiated surface was checked by cross-sectional transmission electron microscopy. For room temperature bombardment a thin amorphous layer (~ 2 nm) was formed on the surface, which was found absent when the experiment was performed at high temperature (300$^0$ C).

## 4. Results and discussion

Figure 1 shows the influence of the incidence angle (θ) on the evolution of the Ge surface topography irradiated with 500 eV Ar$^+$ ions at elevated temperature 300°C and ion fluence $1 \times 10^{19}$ ions cm$^{-2}$. To realize Ge morphological variation due to identical bombardment situation at room temperature, one can see Ref. 18 where we reported parallel mode ripples from θ = 65°-75° and perpendicular mode ripples at 85° respectively. The pattern formation was explained in the framework of curvature dependent ion erosion and smoothening via surface diffusion [35]. In the present case of ion irradiation at target temperature 300$^0$C, we observe four-fold symmetric checkerboard patterns at normal incidence due. The pattern's shape is found rectangular with an orientation in the <100> direction. The roughness increases to about 3.5 nm compared to virgin Ge surface roughness. With increase of ion incidence angle, these structures retain their four-fold symmetry up to 65$^0$ with a slight decrease of the average rms roughness to 2.75 nm. In addition, from 15$^0$ to 35$^0$, the structures are found to hold square shape which again start to show elongation in the beam direction from 45° onwards. The elongation becomes more prominent after θ ≈ 65$^0$ and finally a ripple-like structure appears at θ = 70$^0$. Here, the roughness shows a maximum of 4.25 nm. The wave-vector of the ripples are found to orient along perpendicular direction of ion beam projection and thus, named as perpendicular mode ripples. Beyond 70$^0$, the ripple's regularity gets better and a sharp fall of roughness is observed. With further increase of incidence angle at 85$^0$, the ripples become fainted and the surface become almost smooth again. A further drop of surface roughness is also seen.

For further studies of the order, symmetry and orientation of the nanopattern, we evaluate the fast Fourier transform (FFT) and two-dimensional angle distribution for different angles of incidence from the corresponding AFM images (fig. 1). These are shown at the left side of each AFM images. Both the distributions show a clear signature of facet formation. At low angles, the FFTs in reciprocal space exhibit fourfold symmetric central spot with edges oriented along the <100> crystal directions. Also, the angle distribution shows four distinct peaks which indicate a clear signature of the facet formation tilted away from the high-symmetry orientation.

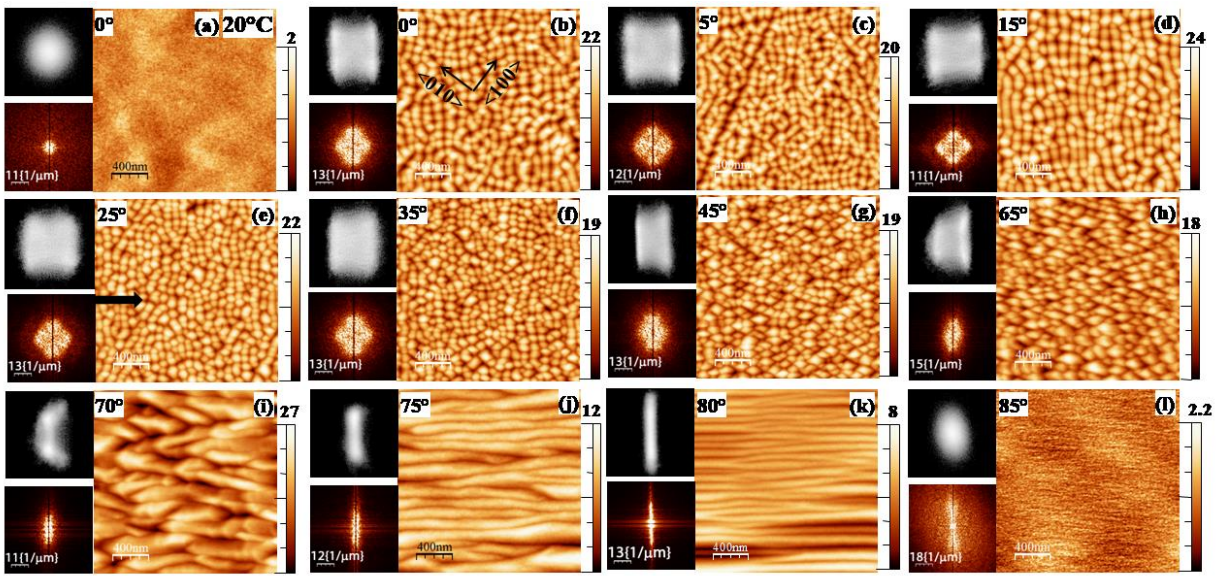

Fig. 1: Ion incidence angle dependence of IBS on Ge surface at $300^0$ C 500 eV $Ar^+ \rightarrow$ Ge ; ion current density = 1000 µA $cm^{-2}$ and ion fluence = $1\times10^{19}$ ions.$cm^{-2}$. The corresponding FFT and 2D angle distribution are shown left side of the AFM images. Black arrow indicates the incident ion beam direction. The left top corner AFM image (a) is at room temperaure irradiation. Height scales are in nm.

As the incident angle increases beyond $45^0$, the 2D-FFT and 2D slope distribution are started to get squeezed in the vertical direction indicating the gradual transformation to two dimensional facet structures (nanoripple) with the orientation parallel to the incident beam direction. Finally at $85^0$ irradiation, although the surface turns out to be flat as revealed from the corresponding AFM image, the central spot of the corresponding FFT shows an elongated structure perpendicular to the direction of ion beam projection thus revealing the reminiscence of the presence of perpendicular mode ripples on the surface.

All the above described observations can be quantitatively visualized from the variation of different surface characteristics, rms roughness ($R_q$), wavelength and facet angle against θ. We calculate the rms roughness or the interface width from the height-height correlation function defined as $G(r) \equiv \langle [h_i - h_j]^2 \rangle$ where $h_i$ and $h_j$ are the heights corresponds to two surface positions separated by a lateral distance $r$. For $r \ll \xi$, $G(r,t)$ scales linearly with $r$ as $r^{2\alpha}$ and for $r \gg \xi$, $\sqrt{G(r)/2}$ gives a measure of the rms roughness, where $\xi$ and $\alpha$ are known as the lateral correlation length and the roughness exponent, respectively. For the patterned structures, calculations of $G(r)$ along the direction of pattern's repetition show a number of oscillations after the linear part of the curve where the first minimum provides the repeat distance between two consecutive height maxima.

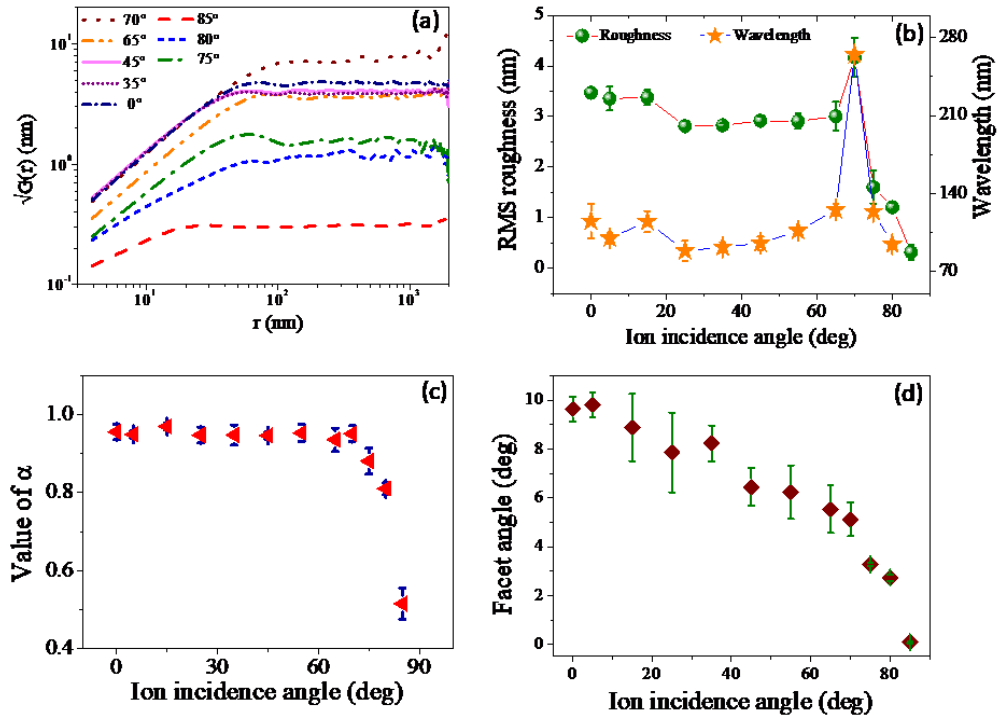

Fig. 2: Ion incidence angle dependent morphological characteristics on Ge(100) surface at implant temperature $300^0$C during $Ar^+$ ion beam sputtering; ion energy = 500 eV, ion current density = 1000 μA $cm^{-2}$ and ion fluence = $1\times10^{19}$ ions.$cm^{-2}$.

In our experiment, the patterns in between $0^0$ to $65^0$ are four fold symmetric mound and pits and beyond that, patterns show repetition in perpendicular direction of the ion beam projection on surface. Thus, we extract $G(r)$ directly from the AFM micrographs assigned for different ion incidence angles along the mentioned direction and plot them in Fig. 2(a) as $\sqrt{G(r,t)}$ versus $r$ in logarithmic scale. Expectedly, the curves show periodic oscillations from $0^0$ to $80^0$ as the surface morphologies consists of the periodic repetition of height modulation. From the first minimum, we extract the wavelength (λ) of patterns and plot in fig. 2(b) as function of incidence angle θ. The overall variation of λ is found to be similar as observed for surface roughness. For small values of θ (up to $15^0$), λ is found around 110 nm which decreases a bit to 95 nm from $25^0$ to $55^0$. Then, λ shows an increase with increase of θ and reaches a maximum of 260 nm at $75^0$. After that, it shows a sudden drop and finally ended with 120 nm at 80°. We also calculate roughness exponent α from the linear part of $\sqrt{G(r,t)}$ versus $r$ curve for each ion incidence angles which are displaed in fig. 2(b). From angular range 0 to $70^0$, α value is found $0.95 \pm 0.01$ which closely corresponds to the theoretically predicted value for a Schwoebel barrier-induced roughening instability [36].

Furthermore, we analyse the facet angles from the two-dimensional slope distributions and its variation with θ is displayed in fig. 2(d). Up to the incidence angle of $35^0$, the facets forms angles about $9^0$ with the (001) surface which plausibly be assigned to plane {106}. Such selection of slope in the continuum description of kinetic instabilities is, again, related to the Schwoebel barrier and is consistent to those observed in Ge homoepitaxy [37-39]. From $45^0$ to $65^0$, where the checkerboard pattern starts showing elongation along the beam direction, the facet angle value drops around $6^0$ which might corresponds to the development of {109} plane. For further increase of incidence angles, $θ \geq 70^0$, the facet angle value is again decreased and finally reaches almost null value at $85^0$ which characterizes a smooth surface and is found in conformity with the roughness variation presented in fig. 2(b).

The above described morphological details indicate that the ES barrier instability is responsible for pattern formation up to incidence angle $65^0$, the existence of which is found only on crystalline surface. In order to ascertain the role of surface crystallinity in pattern formation, we rotate the substrate around its surface normal azimuthally with a constant speed of 5 rpm after fixing it at the respective angles of ion bombardment.

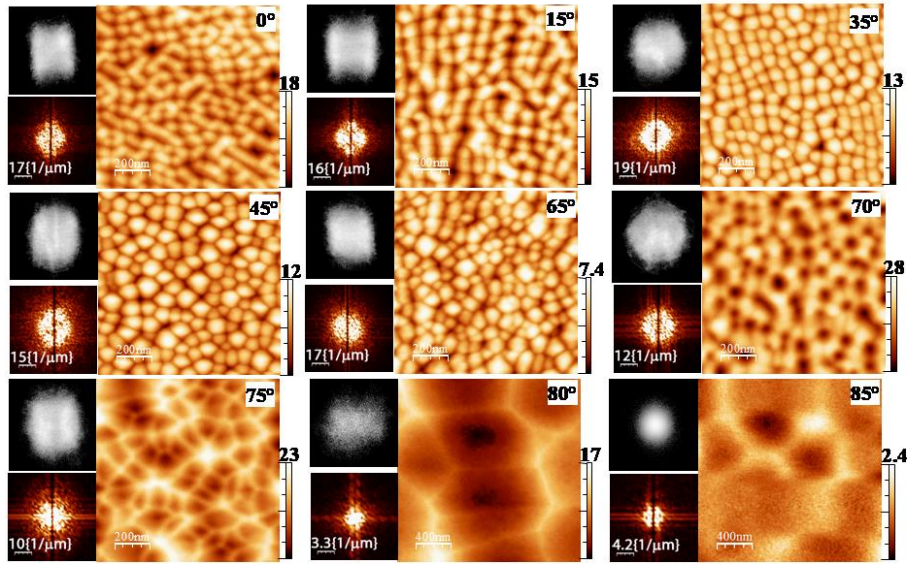

Fig. 3: Result of substrate rotation 5 rpm on the evolution of nanopatterns on Ge (100) surface at different ion incidence angles and temperature $300^0$ C; ion energy = 500 eV, ion current density = 1000 µA cm$^{-2}$ and ion fluence = $1\times10^{19}$ ions.cm$^{-2}$. At the left of the AFM images, the corresponding FFT and 2D angle distribution are shown. Height scales are in nm.

This scenario becomes equivalent to the system where the substrate with a certain tilt is bombarded homogeneously by an ion beam through all azimuthal angles. The results are represented in fig. 3. The AFM images, their corresponding FFT and angle distribution up to $\theta \approx 65^0$, confirms that the symmetry of the pattern remains similar to those obtained for the non-rotating target. This clearly indicates that the four-fold symmetric checkerboard patterns are influenced only by the substrate crystallography i.e., by the ES barrier instability and not by the incident beam direction. However, after $\theta = 65^0$, the results are different; the anisotropic ripple morphology following rotation degenerates into hole/pit structure with feeble reminiscence of four-fold symmetry. Consequently, the corresponding FFT images and the angular distributions show a weak four-fold symmetric structure. At more grazing incidence angle of $80^0$, these pits grow larger in size and at $85^0$, a few of them sustains as isolated large shallow square pits on surface. From the above results, one may say that, beyond the incidence angle $65^0$, pattern formation is dominated by the ion beam direction with respect to the surface normal with weak presence of ES barrier surface instability.

We may, therefore, conclude that two kinds of surface instability play the role in the development of surface topography in the present experimental conditions: (i) the Ehrlich-Schwoebel (ES) barrier induced surface instability at incident angles close to normal, and (ii) the curvature dependent surface instability at grazing incident angles. The ion incidence angle θ ≈ 70⁰ can be thought as the transition angle from the ES barrier instability to curvature dependent ion beam instability for Ge surface at 300⁰C implantation temperature.

We also analyses the morphological characteristics influenced by the substrate rotation at each incidence angles in order to understand the effect of substrate rotation (SR) on surface topographies in contrast to the static/non-rotating (NR) condition.

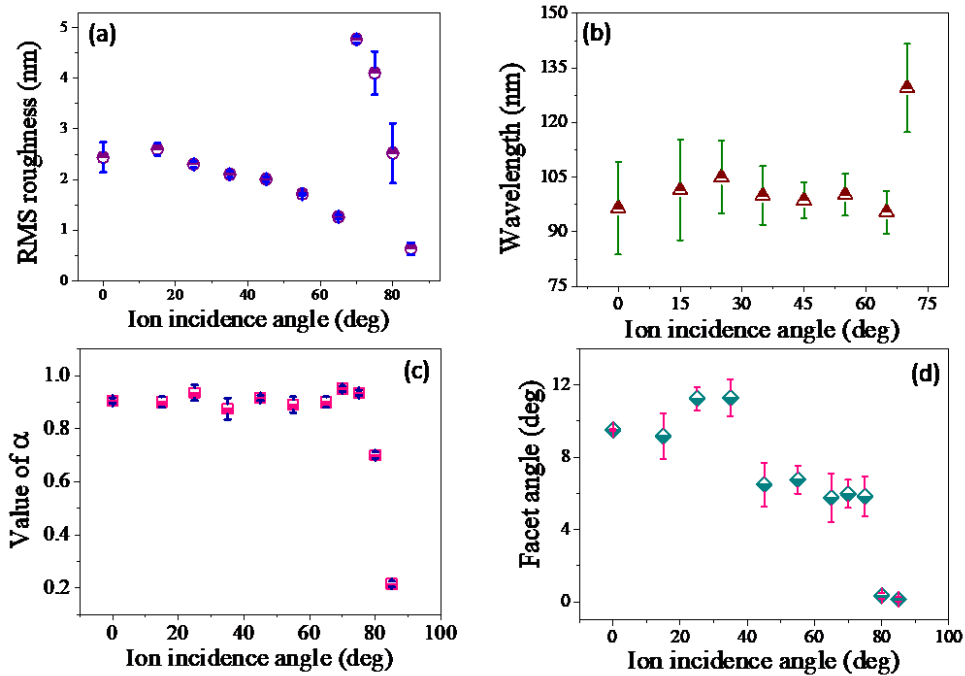

Fig. 4: Variation of (a) RMS roughness, (b) wavelength, (c) roughness exponent α and (d) facet angles of the Ge surface morphologies as function of ion incidence angle with constant substrate rotation 5 rpm at 300⁰ C during Ar⁺ ion beam sputtering; ion energy = 500 eV, ion current density = 1000 μA cm⁻² and ion fluence = 1×10¹⁹ ions.cm⁻².

Surface roughnes $R_{q(SR)}$, wavelength $\lambda_{SR}$ and roughness exponent $\alpha_{SR}$ are determined from the height-height correlation functions (the curves are not shown here) similar to those discussed for NR case and are plotted in fig. 4(a), (b) and (c) respectively as a function of incidence angle $\theta$. We also calculate facet angles from the corresponding slope distributions, which are displayed in fig. 4 (d). The results show some differences with the NR case (cf., fig. 2). The $R_{q(SR)}$ versus $\theta$ shows a bell-shaped type variation with a maximum at $70^0$ similar to $R_{q(NR)}$, but exhibits less roughness amplitude up to $65^0$ and higher afterwards with respect to NR case. On the other hand, the wavelength of the nanostructures $\lambda_{SR}$ remains constant up to $65^0$ followed by a sudden rise at $70^0$ (fig. 4b). This trend looks similar to that observed in NR case (Fig. 2b). However, the $\lambda_{SR}$ values are found lower than the corresponding $\lambda_{NR}$. After $\theta = 70°$, the weakly ordered pits exhibit no more spatial correlation among them which makes the extraction of its wavelength not feasible. Figure 4(c) shows the variation of roughness exponent $\alpha_{SR}$ with $\theta$, which again show similar kind of variation as observed in NR case. The $\alpha_{SR}$ exhibits a value around 0.95 up to $70^0$ followed by a sharp fall with further increase of $\theta$, which might indicate the transition from ES biased instability to curvature dependent surface instability.

In order to further illuminate the pattern formation mechanism, we have investigated the temporal evolution of the checkerboard pattern at oblique incidence angles θ = 15⁰ and 65⁰ and also, of the perpendicular mode ripples at 80⁰. The results are presented in figs. 5(a - e), (f - j) and (k - o) respectively where each of them displayed series of AFM images corresponding to different ion fluences. All the AFM images again enclose their corresponding FFT and two dimensional angle distribution.

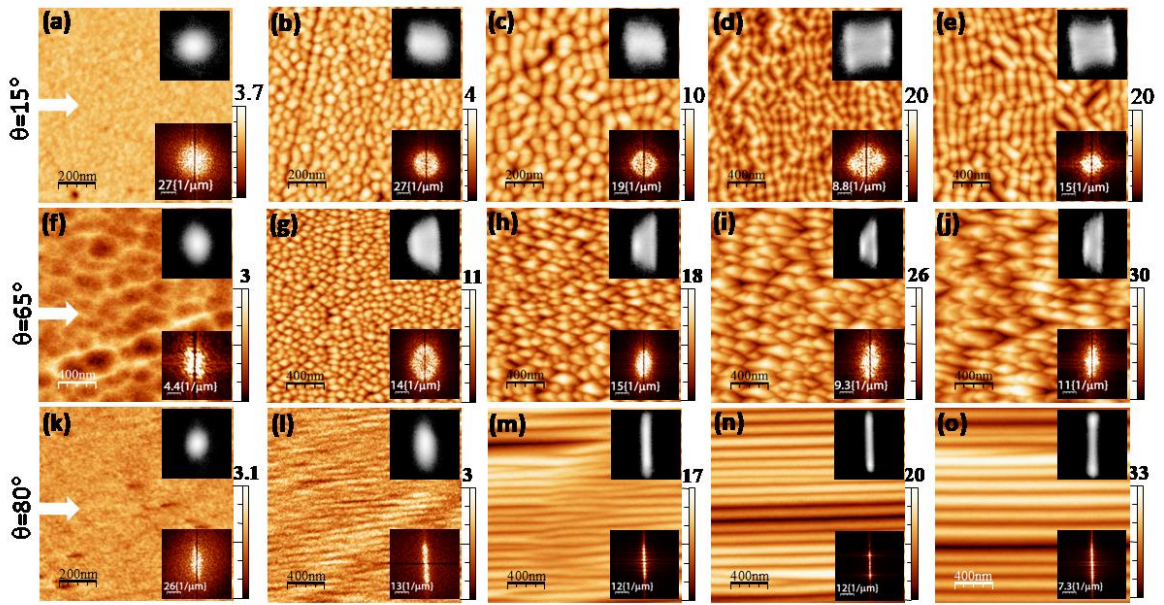

Fig. 5: The AFM topography of the 500 eV Ar⁺ ion irradiated Ge (001) surfaces at 300⁰C and at incidence angles (a-e) θ = 15°, (f-j) 65° and (k-o) 80° for different ion fluences: (a) 1×10¹⁷ions.cm⁻²; (b) 5×10¹⁷ions.cm⁻²; (c) 5×10¹⁸ions.cm⁻²; (d) 2.3×10¹⁹ions.cm⁻² ; (e) 1×10²⁰ions.cm⁻²; (f) 8.6×10¹⁷ions.cm⁻²; (g) 3×10¹⁸ions.cm⁻²;(h) 7.2×10¹⁸ions.cm⁻²; (i) 2×10¹⁹ions.cm⁻²; (j) 5×10¹⁹ions.cm⁻²; (k) 1×10¹⁷ions.cm⁻²; (l) 1.5×10¹⁸ions.cm⁻²; (m) 1.2×10¹⁹ ions.cm⁻²; (n) 2.7×10¹⁹ ions.cm⁻² and (o) 5.3×10¹⁹ions.cm⁻². At the inset of the AFM images, the corresponding FFT images and 2D angle distributions are included. Height scales are in nm. White arrow indicates the incident ion beam direction.

First, the results for $15^0$ will be described and then $65^0$ and $80^0$ will be uncovered in sequence. For ion irradiation at $\theta = 15^0$ and at temperature of $300^0C$, we observe no development of distinctive surface morphology [fig. 5(a)] at initial ion fluence $1\times10^{17}$ ions.cm$^{-2}$. Consequently, the FFT as well as the angle distribution both are found isotropic. The increase of ion fluence to $5\times10^{17}$ ions.cm$^{-2}$ leads a faint topography of mound structures with no clear preference in orientation. Also, the corresponding FFT accordingly displays almost isotropic central spot but the angle distribution starts to demonstrate four-fold symmetric distribution [fig. 5(b)]. A clear four-fold symmetric pattern of alternate mounds and pits with an orientation along <100> crystal direction is found to develop at fluence $1\times10^{19}$ ions.cm$^{-2}$, although the four fold symmetry in FFT becomes more clear from fluence $5\times10^{18}$ ions.cm$^{-2}$. As we move to $\theta = 65^0$, the surface morphology tends to demonstrate random pits at the early fluence regime from $8.6\times10^{17}$ to $2.2\times10^{18}$ ions.cm$^{-2}$ [a representative AFM image is only shown for fluence $8.6\times10^{17}$ ions.cm$^{-2}$ (fig. 5(f))]. Further increase of fluence to $3\times10^{18}$ ions.cm$^{-2}$ leads four-fold symmetric patterns with a slight elongation along the ion beam direction. This behaviour is also confirmed by the corresponding FFT image and the angle distribution as both of them display elongated central spot perpendicular to ion beam direction. The alignment of the patterns along the direction of ion beam gets more pronounced with the increase of ion fluence as seen from the AFM images [figs. 5(h) – (j)]. This characteristic is also reflected in both the FFT and angle distribution showing the increase of squeezing of the central spot along the respective direction. At grazing ion incidence of $80^0$ the AFM image reveals again a smooth surface at lower fluences, although the signature of the perpendicular mode ripple can be confirmed from both the FFT and angle distribution images [fig. 5(k)]. At high ion fluence $1\times10^{19}$ ions.cm$^{-2}$ [fig. 5(m)], the ripples become pronounced but comprises a number of topological defects, e.g., the lateral merging of the two ripple ridges into Y-like junctions. Further increase of ion fluence greatly reduces the number of defects and enhances the ripple ordering.

In order to know quantitatively the morphological characteristics of the evolved patterns at $\theta = 15^0$, $65^0$ and $80^0$, we have extracted the rms roughness ($R_q$), wavelength ($\lambda$) and the facet angle of the patterns from the corresponding AFM images and the data are plotted as a function of ion fluence ($\phi$) in figs. 6 (a) – (c), (d) – (f) and (g) – (i), respectively. For $\theta = 15°$, $R_q$ is found to remain constant up to the fluence $5\times10^{17}$ ions.cm$^{-2}$ [fig. 6(a)]. After which it shows increase with increase of $\phi$ following a power law $R_q \approx \phi^\beta$ with growth exponent $\beta = 0.69 \pm 0.11$ and reaches a saturated value around 3.6 nm at $\phi \geq 1\times10^{19}$ ions.cm$^{-2}$. On the other hand, the wavelength ($\lambda$) and the facet angle, both show an increase from the beginning and saturates at fluences $\phi \geq 1\times10^{19}$ ions.cm$^{-2}$ similar to that observed for the surface roughness. The increment of $\lambda$ from 58 to 120 nm occurs with coarsening exponent

n = 0.24 ± 0.03 ($\lambda \approx \phi^n$), whereas the facet angle initially showing a sharp increase and saturates around $9^0$.

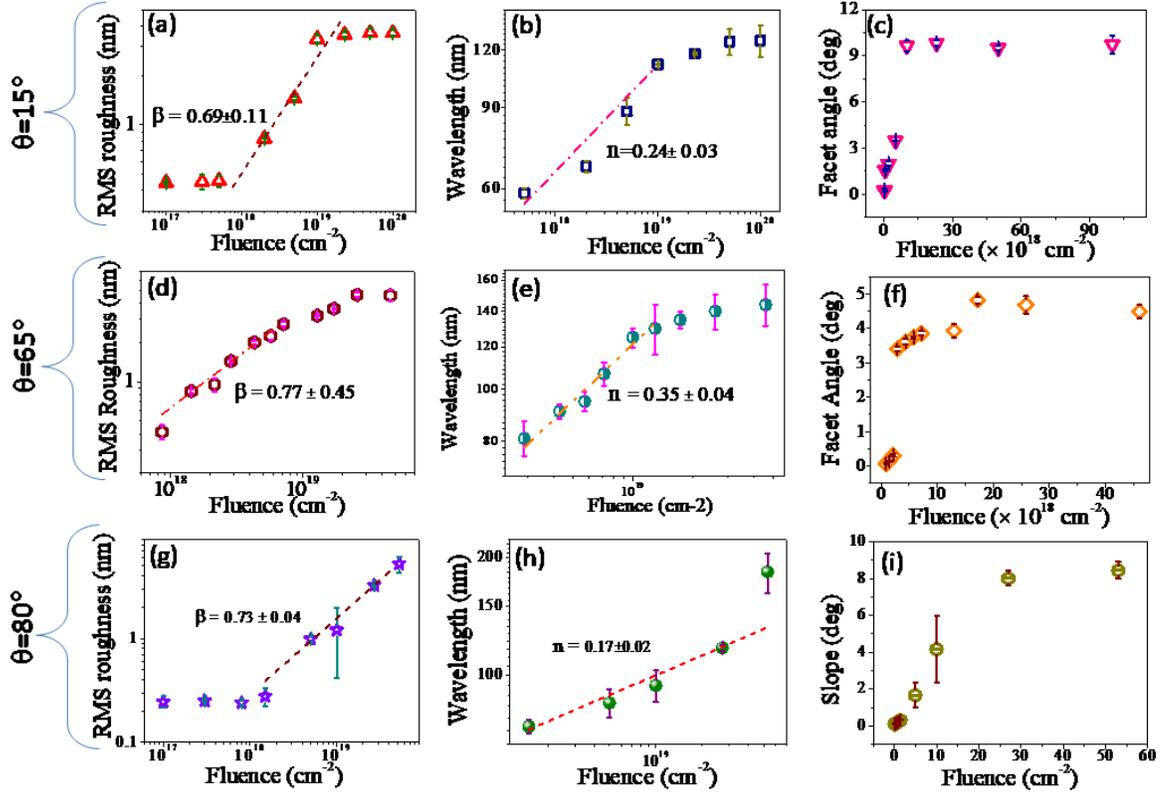

Fig. 6: The RMS roughness, wavelength and facet angle as function of ion fluence for incidence angle 15° (top row: a-c), 65° (middle row: d-f) and 80° (bottom row: g-i) respectively for 500 eV $Ar^+$ ion irradiation on Ge surface at implant temperature 300°C.

For $65^0$ angle of incidence, where the elongated checkerboard patterns are formed, $R_q$, $\lambda$ and the facet angle all of them increase initially and finally saturates at the high fluence $\phi \geq 1 \times 10^{19}$ ions.cm$^{-2}$. Moreover, $R_q$ and $\lambda$ exhibit growth (β) and coarsening (n) exponent as 0.77 ± 0.45 and 0.35 ± 0.04, respectively and their saturated values corresponds to 4.5 nm and 140 nm, respectively. The saturation value for facet angle is found around $5^0$. In case of the evolution of the perpendicular mode ripples at $\theta = 80°$, $R_q$ displays a constant value for initial ion fluences up to $\phi = 1\times10^{19}$ ions.cm$^{-2}$ and then shows increase with growth exponent $\beta = 0.73 \pm 0.04$. On the other hand, the increment of $\lambda$ with ion fluence exhibits the coarsening exponent $0.17 \pm 0.02$. However, no saturation is observed for both $R_q$ and $\lambda$ up to the highest ion fluence of $5\times10^{19}$ ions.cm$^{-2}$. In contrast to $R_q$ and $\lambda$, the facet angle shows saturation around $8°$ after an initial increase with ion fluence.

**Summary**

In summary, we study the pattern formation on Ge surface by low-energy ion irradiation above the dynamic recrystallization temperature 300°C. Up to incidence angle $65^0$, crystalline four-fold symmetric ordered nanostructures are found to develop which resemble the checkerboard patterns observed in case of homoepitaxy. In between, from incidence angle 45°, the patterns show slight elongation along the ion beam projection on surface which becomes prominent at more grazing incidences and finally results highly regular perpendicular mode ripples from 75° onwards. The growth dynamics study reveals that, for initial ion fluences, the amplitude of the patterns grows more rapidly at higher off-normal incidences ($\beta_{65°} \sim 0.77 \pm 0.45 > \beta_{15°} \sim 0.69 \pm 0.11$, where checkerboard patterns develops) below the critical angle of pattern transition $\theta_C \sim 70°$ and after which the growth rate slows down a bit with exponent value $\beta_{80°} \sim 0.73 \pm 0.04$ (where perpendicular mode ripples occur). On the other hand, for high ion fluences, the pattern's amplitude leans towards saturation where the fluence value to reach saturation is found to shift to higher value for larger angle of ion incidence independent of angular regime above or below $\theta_C$. For instance, the evolution of checkerboard patterns at 15° shows saturation at fluence $1 \times 10^{19}$ ion.cm$^{-2}$ whereas the perpendicular mode ripples at 80° do not even saturation even at comparatively five times higher ion fluence value $5 \times 10^{19}$ ion.cm$^{-2}$. Moreover, the checkerboard patterns are found to retain its four-fold symmetry due to concurrent sample rotation during ion irradiation. At the same circumstance, the perpendicular mode ripples completely loses its two fold symmetry and results into pit structures with a weak four-fold symmetry. The overall morphological variation indicates that at low angles of ion irradiation, the kinetic instability is largely influenced by the surface diffusion of vacancies due to presence of ES barrier, whereas at grazing incidence angles, the surface curvature dependent ion beam induced instability dominates.

## Acknowledgement

One of the authors (DG) as Emeritus Scientist, CSIR, is thankful to CSIR, New Delhi (Grant No. 21(0988)/13/EMR-II dated 30-04-2015) for providing financial support.